\documentclass[12pt]{iopart}

\usepackage[]{graphicx}

\begin{document}

\title[Parametric amplification with weak-link nonlinearity]{Parametric amplification with weak-link nonlinearity in superconducting microresonators}

\author{E A Thol\'en$^1$, A Erg\"ul$^1$, K Stannigel$^2$\footnote{Present address: Institute for Quantum Optics and Quantum Information of the Austrian Academy of Sciences, A-6020 Innsbruck, Austria}, C Hutter$^2$ and D~B~Haviland$^1$}

\address{$^1$ Nanostructure Physics, Royal Institute of Technology (KTH), Albanova, SE-10791 Stockholm, Sweden.}

\address{$^2$ Department of Physics, Stockholm University, AlbaNova University Center, SE-10691 Stockholm, Sweden}

\ead{haviland@kth.se}

\begin{abstract}
Nonlinear kinetic inductance in a high $Q$ superconducting coplanar waveguide microresonator can cause a bifurcation of the resonance curve. Near the critical pumping power and frequency for bifurcation, large parametric gain is observed for signals in the frequency band near resonance. We show experimental results on signal and intermodulation gain which are well described by a theory of the parametric amplification based on a Kerr nonlinearity. Phase dependent gain, or signal squeezing, is verified with a homodyne detection scheme.
\end{abstract}

\submitto{\PS}

\section{Introduction}

Interest in microwave parametric amplifiers is being renewed by the need for low noise, minimal back-action measurement in the study of quantum electrodynamics with superconducting circuits \cite{clerk:reviewquantumnoise:08}. Work on microwave parametric amplifiers dates back to the early days of high frequency electronics and the development of radar \cite{louisell:electronics:60}. Today most microwave amplifiers are based on discrete components and feedback, employing high electron mobility transistors. These amplifiers, operating at cryogenic temperatures, are achieving ever higher frequencies and ever lower noise figures, with high gain in a broad frequency band, and they are indispensable as second-stage amplifiers for the experiments described here. However, parametric amplifiers of the type discussed in this work are able to achieve even lower noise figures, and they possess unique qualities which motivate their further development as a general purpose, quantum-limited measurement tool. Here we study a parametric amplifier based on nonlinear kinetic inductance of a weak link in a superconducting resonator \cite{abdo:intermod-gain-nbn:06, segev:selfsustained:07}. Parametric amplifiers based on the nonlinear inductance of a Josephson junction have been previously explored for their low noise properties \cite{Wahlsten:JJarrayParaAmp:78, yurke:joseph-par-amp:89, kuzmin:paraamp:85}.

A parametric amplifier does not amplify signals in the conventional sense. While it does provide power gain $G$, it also ``squeezes'' a signal. If we describe an arbitrary signal at the input of the parametric amplifier by its two quadrature components, $A(t) = A_1\cos{\omega t} + A_2\sin{\omega t}$, the quadrature signals at the output of the parametric amplifier can be written as $B(t)=\sqrt{G}A_1 \cos{(\omega t + \phi)} + \left(1/\sqrt{G}\right) A_2 \sin{(\omega t + \phi)}$. Thus one quadrature is amplified, while the other is deamplified. This squeezing is, in principle, a reversible process -- it can be undone by the application of a second parametric amplifier with its internal clock appropriately phase-shifted with respect to the first amplifiers clock. The reversibility of this amplification process means that the amplification can occur without adding any noise to a signal. Thus, the parametric amplifier can amplify a signal at the quantum limit, delivering gain to a signal while the signal remains coherent, with the minimal uncertainty that is imposed by quantum physics,
\begin{equation}
 \Delta B_1 \Delta B_2 = \sqrt{G} \Delta A_1 \frac{1}{\sqrt{G}}\Delta A_2 = \frac{1}{2} \hbar \omega Z f_\mathrm{B},
\end{equation}
where $Z$ is the characteristic impedance of the signal source and load which are assumed identical and matched to the amplifier, and $f_\mathrm{B}$ is the frequency bandwidth over which we measure gain. A phase-insensitive amplifier, which amplifies both quadratures with the same gain $\sqrt{G}$, must add at least $(G-1)\frac{1}{2} \hbar \omega f_\mathrm{B} $ of noise power to the output signal \cite{caves:quantumlimitedamps:82, yurke:backactionevasion:91}.

\begin{figure}
 \centering
 \includegraphics{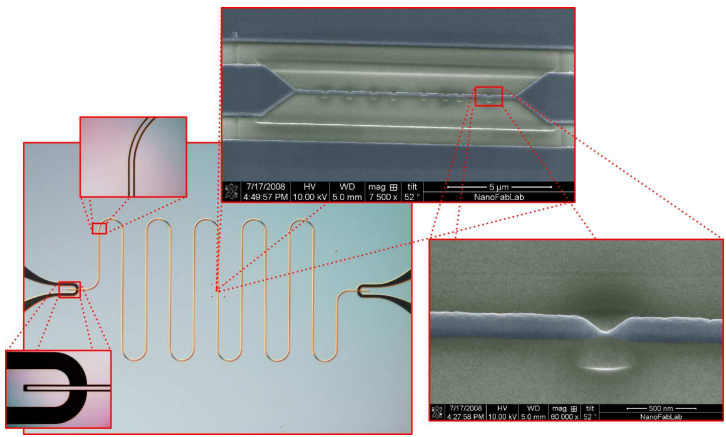}
 \caption{CPW resonators were made of Nb films using photo-lithography. The nonlinear inductance is due to a weak link in the center of the CPW resonator, where a focused ion beam was used to make a narrow channel with 10 small notches.}
 \label{sample}
\end{figure}

Parametric amplification can occur near the bifurcation of a nonlinear oscillator. The nonlinear oscillator used here is a standing wave mode of a superconducting coplanar waveguide (CPW) resonator \cite{tholen:paraamp_apl:07}, with an intentional weak-link at the anti-node of the current (i.e.\ in the center of the resonator), see figure~\ref{sample}. Superconducting CPW resonators have very low internal loss and their small transverse dimensions in comparison with the modal wavelength give a small modal volume, which is advantageous for strong coupling of the resonator modes to quantum circuits \cite{blais:cavity-QED:04,wallraff:cavity:04,frunzio:cavity-fab:05}. These microwave resonators have been instrumental in the development of circuit QED, where they are often used in the linear regime. They can be used for example to probe the quantum state of a Cooper pair box by measuring a shift of the oscillator's resonance \cite{wallraff:visibility:05}. Recently, CPW resonators with nonlinear Josephson junction elements have been used to realize bifurcation amplifiers \cite{metcalfe:MeasuringQubitJBA:07, boulant:qnd_bifurcationamp:07, bertet:thisproc:09} and parametric amplifiers \cite{castellanos:noisesqueezing:08,castellanos:paraamp:07, bergeal:jjringparaamp:08, abdo:squidParaamp:09, wilson:thisproc:09}.

\begin{figure}
 \centering
 \includegraphics{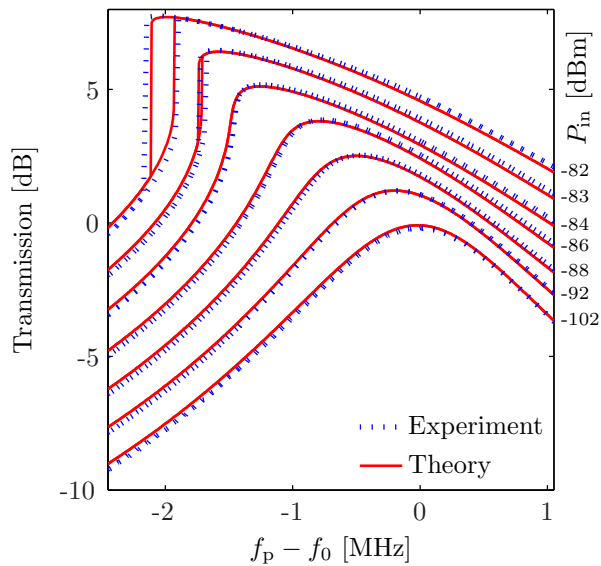}
 \caption{ Bending of the resonance curve, transmission vs.\ frequency, of a high $Q$ superconducting coplanar waveguide resonator. Each curve is offset for clarity. From bottom to top the drive power is increased as indicated. The dotted lines show the measured data and the solid lines show fits to the data with the nonlinear model described in the text.}
 \label{bifurcation}
\end{figure}

The weak damping, or high quality factor $Q \approx 10^3 - 10^6$, of superconducting CPW resonators means that they build up large amplitude oscillations near resonance, even for a very weak driving signal. If the resonator is slightly anharmonic (nonlinear restoring force) there will be a bending of the resonance curve as the drive power is increased, which eventually results in hysteretic behavior of the resonance curve (see figure~\ref{bifurcation}). Bifurcation occurs at a critical power and frequency, right at the onset of the double-valued resonance curve. At this critical point the oscillator dynamics suddenly develops qualitatively new behavior, changing from one stable state of oscillation to three states of oscillation, two of which are stable and one which is unstable. When the driven, damped nonlinear oscillator is balanced in the unstable oscillation state, a small perturbation will cause the oscillator to evolve toward one of the two stable oscillation states. This sensitivity of the bifurcated oscillator is exploited using a pulse-and-hold measurement technique \cite{walter:PulseHold:07} in the bifurcation amplifier \cite{siddiqi:bifurcation:04}. The bifurcation amplifier can measure the state of a qubit with high fidelity \cite{metcalfe:MeasuringQubitJBA:07, boulant:qnd_bifurcationamp:07, bertet:thisproc:09,lupascu:dispersive:06}, giving a digital output, where the result of measurement is that the oscillator ends up in one of its two possible stable states. The parametric amplifier also exploits the dynamic instability near bifurcation, but in an analog way, with a continuous drive called the pump. 

Parametric amplifiers are often operated in what is known as ``degenerate mode'', where the nonlinear oscillator is driven, or pumped, at twice the resonance frequency. When the pump amplitude is large enough, a period-doubling bifurcation can occur. Period-doubling means that response will spontaneously develop at one-half the pump frequency, on resonance. When the pump is tuned slightly below the critical power for the onset of period-doubling bifurcation, a weak signal tone which is injected at one half the pump frequency will be squeezed. The pump, being at exactly twice the signal frequency, has a fixed phase relation to the signal. Signals in phase with the pump are amplified, whereas signals out of phase with the pump are deamplified. Gain is achieved if the signal and pump have the correct phase relation. By gain, we mean that the nonlinear oscillator driven in this way can deliver more power to a load at the signal frequency, than was supplied at the signal frequency. The nonlinearity causes power to be taken from the pump tone, and transfered to the signal tone, something that is not possible for linear systems, where energy in each harmonic tone is strictly conserved. 

In this work we operate the parametric amplifier in the so-called ``non-degenerate mode'', where the pump is near resonance, very close to the critical frequency and power for bifurcation of the nonlinear oscillator, see figure~\ref{bifurcation}. A weak signal tone which is placed close to the pump tone (within the bandwidth of the oscillator) will be stimulated by the pump and amplified. In the non-degenerate mode of operation, the signal and pump no longer have constant phase relation to one another. When the signal and pump are separated by $\Delta f = f_\mathrm{s} - f_\mathrm{p} $, their relative phase is changing periodically with the beat frequency $\Delta f$. However, intermodulation of the signal and the pump tones in the nonlinear oscillator generate response at a third, idler frequency $f_\mathrm{i} = f_\mathrm{p} - \Delta f$. The signal and the idler are both beating against the pump with frequency $\Delta f$, and the correlation of the phases of the signal and its idler, which are both locked to the pump, gives rise to squeezing. 

\begin{figure}
 \centering
 \includegraphics{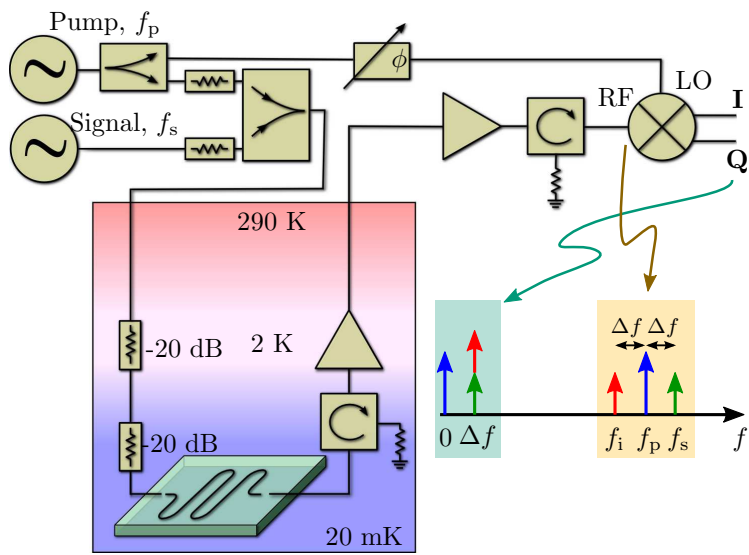}
 \caption{An illustration of the microwave circuitry used to realize the non-degenerate parametric amplifier. A strong pump at $f_\mathrm{p}$ and a weak signal at $f_\mathrm{s}$ are combined and applied to the nonlinear resonator, which generates an idler at $f_\mathrm{i}$. The homodyne detection scheme mixes the output from the resonator with a phase shifted copy of the pump. The mixer down-converts $f_\mathrm{s}$ and $f_\mathrm{i}$ to the same frequency $\Delta f$. Squeezing is revealed as a $\phi$ dependence of the response at $\Delta f$ resulting from a correlation in the phases signal and its idler.}
 \label{schematic}
\end{figure}

Thus, unlike the degenerate parametric amplifier, the non-degenerate parametric amplifier does not require a particular phase at the input in order to amplify a signal. Any phase of the input signal will be amplified equally well because the amplifier automatically generates the appropriate idler signal. The ability of the non-degenerate parametric amplifier to squeeze can only be revealed after mixing the nonlinear oscillators response, including the signal and its idler, with a phase-shifted pump signal as shown in figure~\ref{schematic}. This mixing causes the signal tone and the idler tone to be down-converted to the same frequency $\Delta f$. The amplitude of the down-converted signal at $\Delta f$ will depend on the phase shift $\phi$.

\section{Experiment and Analysis}
Superconducting CPW resonators were fabricated by STAR Cryoelectronics \cite{STARCRYO} from Nb films by etching out narrow gaps between the center conductor, ground planes, and input/output couplers. We used a focused ion beam to form a thin wire, 10 $\mu$m long with cross section $200\times200$~nm$^2$, in the center of the cavity. In this thin wire 10 constrictions of cross section $\sim 30\times30$~nm$^2$ were formed, see figure~\ref{sample}. 

Measurements were made with the resonator mounted in a dip-stick style dilution refrigerator with base temperature of 20~mK. A schematic of our microwave circuit is shown in figure~\ref{schematic}. Two stages of cold attenuators were used at the input side of the resonator to bring the noise at the input into thermal equilibrium at the base temperature. One circulator was used at the output to shield the resonator from back-action noise due to the second stage amplifier. As a second stage amplifier, we have tested cryogenic amplifiers from Miteq and Low Noise Factory \cite{LNF}, the latter being far superior to the former, with a gain of 38-40~dB in the frequency band 4-8~GHz, a noise temperature of 2-3~K and a DC power consumption of 7~mW. 

The nonlinear CPW resonators are initially characterized by a transmission measurement, without the pump and mixer. To characterize the nonlinearity, we sweep the frequency through resonance and back again, tracing out the resonance curve for different drive powers. The result is shown in figure~\ref{bifurcation}. We observe a bending of the resonance curve toward {\em lower} frequency as the drive power is increased. The shifting of the resonance frequency $\omega_0=1/\sqrt{LC}$ toward lower frequency means that the inductance $L$ of the resonator is increasing with increasing current. Since kinetic inductance is proportional to the inverse of the density of superconducting electrons $L_\mathrm{k} \sim 1/n_\mathrm{s}$, it follows that a reduction of $n_\mathrm{s}$ or pair breaking is the source of nonlinearity.  Indeed, if we envision the weak links as narrow constrictions of a bulk superconductor, then we can describe pair breaking as a consequence of the finite superfluid velocity, within the context of the Ginzburg-Landau theory of superconductivity \cite{tinkham:introsupercon:80}.  Alternatively we could consider the nonlinear current-phase relation $I(\delta)$ of the weak link as the source of nonlinear inductance, $L^{-1}=\frac{2e}{\hbar}\frac{dI} {d\delta}$. The exact form of $I(\delta)$ will depend on the nature of the weak link \cite{likharev:weaklinks:revmodphys1979, chauvin:atomiccontacts:2005}).  In any case, an expansion of this current-phase relation to the lowest nonlinear order in phase (i.e.\ $\delta^3$ term) results in a nonlinear oscillator equation for the resonator which is the Duffing equation.  If we compare the two extreme cases of  a perfectly clean weak link comprised of normal conduction channels of unity transparency, $I(\delta)=I_\mathrm{C} \sin(\delta)/|\cos(\frac{\delta}{2})|$, with that of a dirty weak link were all channels have transparency much less than one, $I(\delta)=I_\mathrm{C}\sin(\delta)$, we find that for the same critical current $I_\mathrm{C}$, the $\delta^3$ term is smaller in the former case by a factor 4.  Thus, clean weak links have a weaker nonlinearity than dirty weak links or tunnel junctions, but in all cases, the scale of the linear and nonlinear inductance is set by $I_\mathrm{C}^{-1}$.  In our experiment we can not independently measure $I_\mathrm{C}$ or $I(\delta)$ but we can determine the amplitude of the standing current wave when the nonlinear oscillator bifurcates, which was $I_\mathrm{b} =19\mu\textrm{A} \ll I_\mathrm{C}$ for the weak link resonator described here.
  
A theory of the parametric amplifier based on a nonlinear oscillator has been recently worked out \cite{yurke:kerr-performance:06, babourina:qnoiseduffing:08} starting from the Hamiltonian: \mbox{$H_0=\hbar\omega_0A^\dagger A + \frac{1}{2}\hbar K A^\dagger A^\dagger A A$}, where the Kerr constant $K$ gives the strength of the nonlinearity. Treated classically, this is the Hamiltonian of the Duffing oscillator. The theory of Yurke and Buks treats the quantum case, where the methods of Gardiner and Collett \cite{gardiner:inputoutput:85} are used to model linear damping due to the input/output port, and a port which models nonlinear damping. The theory calculates response at the signal and idler frequencies and neglects intermodulation products generated at other frequencies. 

In figure~\ref{bifurcation} we see that the transmission data are well described by the theory over a wide range of drive power. By fitting the theory to the data, we are able to extract the following parameters which characterize the nonlinear oscillator: The resonance frequency $\omega_0/2\pi=$7.60845~GHz, the Kerr constant $K/\omega_0=-1.27\times10^{-9}$, the linear damping coefficients due to radiation to the input and output ports $\gamma_1/2\pi=\gamma_4/2\pi=464$~kHz, and the linear damping coefficient due to intrinsic losses in the cavity $\gamma_2/2\pi=9$~kHz. Damping is clearly dominated by radiation losses in this over-coupled cavity, with a total quality factor $Q=\omega_0/2(\gamma_1+\gamma_2+\gamma_4)=4060$. We achieve excellent quantitative agreement between the measured transmission data and the theory without nonlinear damping.

\begin{figure}
 \centering
 \includegraphics{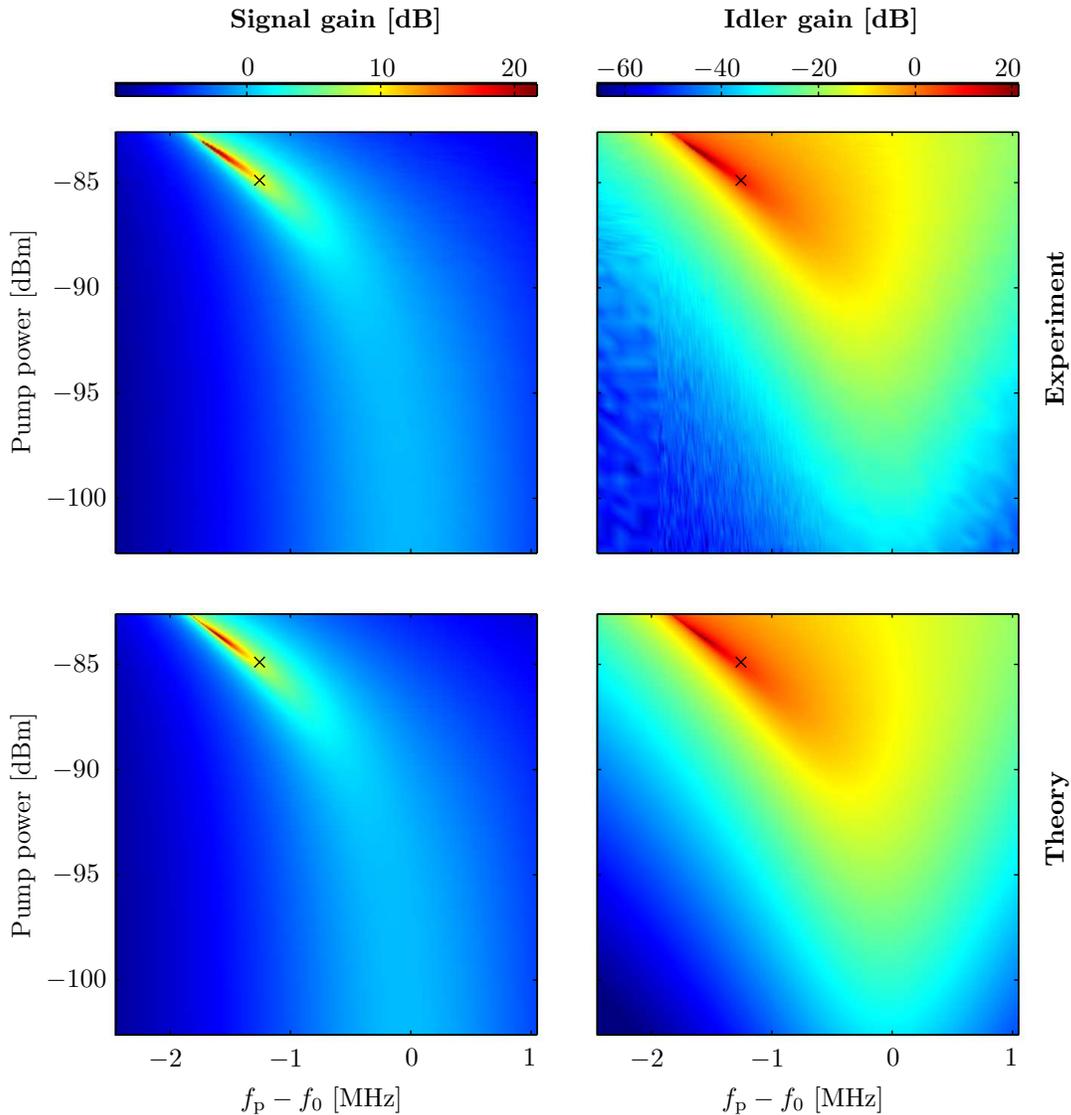}
 \caption{Measured (upper panels) and predicted (lower panels) gain for the signal (left) and idler (right) plotted with a colour map as a function of pump power and pump frequency. At every point the signal frequency is $\Delta f=$~10~kHz above the pump frequency. The cross marks the operating point for the squeezing experiment (see figure~\ref{phase}).}
 \label{exp_the}
\end{figure}

The nonlinear oscillator is operated as a parametric amplifier by combining the pump and signal, and applying this combined signal to the input port of the resonator (see figure~\ref{schematic}). Response is measured at the output port at the signal, pump and idler frequencies. In figure~\ref{exp_the} we see the experimental signal and idler response plotted with a colour map (upper panels), as a function of frequency and pump power. At each pump power we measured transmission at all three frequencies, while sweeping both $f_\mathrm{s}$ and $f_\mathrm{p}$, keeping the spacing $\Delta f = f_\mathrm{s} - f_\mathrm{p}$ constant. Successive sweeps are made with constant signal amplitude, incrementing the pump power. We can see that for the lowest pump power, the signal is unaffected by the pump, and the response is the Lorentzian of a linear oscillator, with nearly unity transmission on resonance. As the pump power is increased and the resonance shifts to lower frequencies, the signal response begins to show gain, reaching a maximum value that exceeds the applied level. When the pump power and frequency are at the threshold for bifurcation of the nonlinear oscillator, the signal response has a sharp peak providing a maximum gain of 22~dB. In contrast to the signal, the idler shows no response at low pump power, as expected when the oscillator is in the linear regime. For increasing pump power we see the idler response emerge. It is interesting to note that the intermodulation of the signal and pump, which produces the idler, is a very sensitive method for detecting the presence of nonlinearity in the oscillator. At very low pump power ($-100$~dBm), where there is no detectable bending of the resonance curve, we can easily distinguish the presence of an idler response above the background noise level. With increasing pump power, the idler response becomes more pronounced, and at the maximum of signal gain, the idler and signal have reached nearly the same amplitude.

Theoretical expressions for the signal and idler response as a function of pump power and frequency are given in ref. \cite{yurke:kerr-performance:06}. We have modified these expressions for our experiments, where two ports are used and transmission rather than reflection is measured \cite{stannigel:mastersthesis:07}. In figure~\ref{exp_the} we plot the theoretical predictions (lower panels) for the parameters given above, as determined from the fits shown in figure~\ref{bifurcation}. We see excellent agreement between the gain predicted by the theory and the measured gain, with no adjustment of the parameters. 

\begin{figure}
 \centering
 \includegraphics{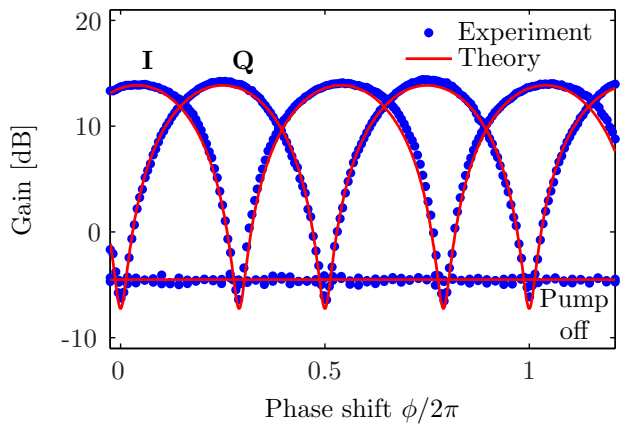}
 \caption{Phase dependence of the down-converted signals \textbf{I} and \textbf{Q} from the mixer (see figure~\ref{schematic}).}
 \label{phase}
\end{figure}

Thus our nonlinear resonator functions as a non-degenerate parametric amplifier with quite reasonable gain (peak gain 25~dB), albeit in a very narrow frequency band (3~dB bandwidth 30 kHz). To demonstrate the phase sensitive nature of the non-degenerate parametric amplifier and to reveal squeezing, we mix the output, containing the pump, signal and idler frequencies, with a phase-shifted copy of the pump signal as shown in figure~\ref{schematic}. In this homodyne scheme, the mixer translates the frequencies inserted to its RF port by an amount $f_\mathrm{p}$, see figure~\ref{schematic}, and divides the translated signals into two quadratures which are delivered at the \textbf{I} and \textbf{Q} ports. In figure~\ref{phase} we show the measured \textbf{I} and \textbf{Q} response at the frequency $\Delta f$, which contains both signal and idler components. The interference between the signal and idler components which depends on the phase-shift $\phi$ gives rise to the squeezing. Here we also see excellent agreement between experiment and the theoretical predictions for squeezing, where we have compensated for the non-ideality of our mixer by adjusting the relative phase of the \textbf{I} and \textbf{Q} signals slightly away from $\pi/2$. For this operating point (marked in figure \ref{exp_the}) we are able to observe $-6.5$~dB of squeezing measured relative to the applied signal level, which corresponds to $-2$~dB relative to the transmission level measured with the pump off.

\section {Discussion and summary}
While the parametric amplifier had relatively good gain, we did not observe ideal behaviour in the sense that the squeezing $1/G$ was not equal to the inverse of the gain $G$. Nevertheless, our results are consistent with the theory for the operation point used in the measurement shown in figure~\ref{phase}. We found that the theory did predict more ideal behavior when the working point was adjusted closer to, but slightly away from, the point of peak gain. However, when performing the experiment near this point we found that the amplifier became unstable. 

The theory of Yurke and Buks \cite{yurke:kerr-performance:06} includes a nonlinear loss parameter $\gamma_3$. For the range of applied power described in this paper, from $-110$~dBm up to $-80$~dBm, which includes the bifurcation at $-83.5$~dBm, we are able to explain our data well with $\gamma_3 = 0$ and a single Kerr nonlinearity $K$. However, for higher power levels, above $-80$~dBm, we find that it is necessary to include $\gamma_3$ and additional expansion coefficients of the nonlinear inductance in order to explain the measured distortion of the resonance curves. Furthermore, when increasing the signal level in the parametric amplifier, we observe intermodulation response at frequencies $2\Delta f$, $3\Delta f$ etc.\ away from the pump. To explain this response one must consider either higher order coefficients in the expansion of the nonlinear inductance, or higher orders of perturbation theory in the response calculated with a Kerr nonlinearity, or both. These higher order effects may become important near the point where the gain peaks, and may be the cause of instability.

In summary, we have demonstrated how a superconducting CPW resonator with a weak link can be used to realize a non-degenerate parametric amplifier. The amplifier exhibits a gain of 22~dB for a signal 10 kHz away from the pump. The gain is sharply peaked in a narrow band centered around the bifurcation point of the nonlinear oscillator. Squeezing of the signal was shown with a homodyne detection scheme, but instability prevented operation near the optimal point. When stable operation was achieved, the measured results are in excellent agreement with a theory \cite{yurke:kerr-performance:06} based on Kerr nonlinearity, with insignificant nonlinear loss. Work is in progress to explore the noise properties of this amplifier.

\ack
We acknowledge stimulating discussions with Michel Devoret, Steve Girvin, Rob Schoelkopf, Konrad Lehnert and Manuel Castellanos-Beltran. Support for this research comes from the Swedish VR and the EU project SCOPE under FET-Open grant number 218783.

\section*{References}

\bibliographystyle{unsrt}
\bibliography{references}

\end{document}